# Universality of Dzyaloshinskii-Moriya interaction effect over domain-wall creep and flow regimes

Duck-Ho Kim,[1] Sang-Cheol Yoo,[1,2] Dae-Yun Kim,[1] Byoung-Chul Min,[2] and Sug-Bong Choe[1†]

[1]Department of Physics and Institute of Applied Physics, Seoul National University, Seoul, 08826, Republic of Korea.

[2]Center for Spintronics, Korea Institute of Science and Technology, Seoul, 02792, Republic of Korea.

†Correspondence to: sugbong@snu.ac.kr

Chirality causes diverse phenomena in nature such as the formation of biological molecules[1], antimatters[2], non-collinear spin structures[3], and magnetic skyrmions[4]. The chirality in magnetic materials is often caused by the noncollinear exchange interaction, called the Dzyaloshinskii-Moriya interaction (DMI)[5,6]. The DMI produces topological spin alignments such as the magnetic skyrmions[4] and chiral domain walls (DWs)[7-12]. In the chiral DWs, the DMI generates an effective magnetic field $H_{DMI}$[13], resulting in a peculiar DW speed variation in the DW creep regime[14-16]. However, the role of $H_{DMI}$ over the different DW-dynamics regimes remains elusive, particularly due to recent observation of distinct behaviors between the creep and flow regimes[17]. We hereby demonstrate experimentally that the role of $H_{DMI}$ is invariant over the creep and flow regimes. In the experiments, the pure DMI effect is quantified by decomposing the symmetric and antisymmetric contributions of the DW motion. The results manifest **that the antisymmetric contribution vanishes gradually across the creep and flow regimes, revealing that the symmetric contribution from $H_{\text{DMI}}$ is unchanged. Though the DW dynamics is governed by distinct mechanisms, the present observation demonstrates the uniqueness of the DMI effect on the DWs over the creep and flow regimes.**

The DMI[5,6] has recently shed new light on prominence due to recent findings on the dynamics of magnetic chiral DWs[7-12]. Up to recent, Je *et al.*[14] demonstrated that, in the DW creep regime, $H_{\text{DMI}}$ modifies the DW energy density and causes variation of the DW speed under the influence of an in-plane magnetic field $H_{\text{in}}$. Such $H_{\text{in}}$-dependence of the DW speed is found to be symmetric for inversion with respect to $H_{\text{DMI}}$; thus, one can quantify the sign and magnitude of $H_{\text{DMI}}$ by symmetry measurement of the DW speed with respect to $H_{\text{in}}$[14,15]. However, recently Jué *et al.*[18] proposed that energy dissipation—called chiral damping—generates additional variation of the DW speed, which is antisymmetric for inversion with respect to $H_{\text{DMI}}$. Because of such sizeable antisymmetric contribution, the symmetry-based $H_{\text{DMI}}$ determination becomes controversial in the DW creep regime. To avoid this controversial issue, Vaňatka *et al.*[17] demonstrated that in the flow regime, symmetry-based $H_{\text{DMI}}$ determination becomes possible, as the flow regime exhibits symmetric DW speed variation, possibly due to the formation of soliton-like Bloch-type DWs above the Walker breakdown[19].

By developing another experimental scheme, we decompose the symmetric and antisymmetric contributions in the creep regime. This scheme is based on the fact that the symmetric and antisymmetric contributions exhibit distinct dependences on an out-of-plane magnetic field. The experimental results clearly show that the antisymmetric contribution gradually vanishes across the creep and flow regimes, while the symmetric contribution remains unchanged, confirming the uniqueness of $H_{\text{DMI}}$ across the creep and flow regimes.

For this study, we chose with weak perpendicular magnetic anisotropy (PMA) films because of their high mobility of field-driven DW motion[20]. The field-driven DW speed $v$ was then measured over a wide range from $5\times10^{-4}$ to 20 m/s across the creep and flow regimes. Figure 1a plots $v$ as a function of out-of-plane magnetic field $H_z$ with in-plane magnetic field $H_x = 0$ mT. The black arrow in the figure indicates the DW depinning field $H_{\text{dep}}$, above which the DW exhibits a dissipative viscous motion with $v \propto H_z - H_{\text{dep}}$, as guided by the red line of the best linear fit. Therefore, this DW motion belongs to the flow regime ($H_z > H_{\text{dep}}$). On the other hand, the creep regime ($H_z < H_{\text{dep}}$) exhibits thermally activated DW motion with the creep criticality $\ln(v) \propto H_z^{-1/4}$, as guided by the blue line of the best linear fit in Fig. 1b.

The effect of $H_x$ on the DW motion was then examined. Figures 2a-d show $v/v_{\min}$ with respect to $H_x$ under various strengths of $H_z$, i.e., 1.9, 3.4, 17, and 41 mT, respectively, over the creep and flow regimes. Here, $v_{\min}$ is the apparent minimum of $v$ as defined below. The best parabolic fitting (solid curve) is shown in each plot to guide the symmetry of $v$, indicating the in-plane magnetic field $H_{\min}$ (purple arrows) for $v_{\min}$ (i.e., $v_{\min} \equiv v(H_{\min})$). It is interesting to note that $H_{\min}$ is shifted across the plots with respect to the strength of $H_z$. Therefore, $H_{\min}$ of the creep regime (green dotted line) differs from that of the flow regime (blue dotted line). As the $H_{\text{DMI}}$-determination scheme is based on the measurement of the inversion symmetry with respect to $H_{\min}$, this observation indicates that $H_{\text{DMI}}$ cannot be uniquely determined irrespective of $H_z$. Because of better symmetries of $v(H_x)$ observed in the flow regime, Vaňatka et al.[17] have argued that $H_{\min}$ measured in the flow regime truly quantifies $H_{\text{DMI}}$ ($= -H_{\min}$), whereas the asymmetric behavior in the creep regime contains sizeable antisymmetric contribution.

To confirm whether $H_{\min}$ measured in the flow regime corresponds to the true $H_{\mathrm{DMI}}$, we further analyze the DW motion in the creep regime by decomposing the symmetric and antisymmetric contributions. The principle of the decomposition is as follows: recent studies[14,21,22] have proposed that in the creep regime, the thermally activated DW motion follows the DW creep criticality[23] as given by $v(H_x, H_z) = v_0(H_x) \exp[-\alpha(H_x) H_z^{-1/4}]$, where $v_0$ is the characteristic speed and $\alpha$ is a scaling parameter related to the energy. According to ref.14, $\alpha(H_x)$ mainly attributes the symmetric contribution by varying the DW energy density. On the other hand, it has been proposed in ref.18 that $v_0(H_x)$ possibly includes a sizeable antisymmetric contribution, even though the nature of $v_0(H_x)$ is not fully understood yet. The troublesome $v_0(H_x)$ can be easily removed experimentally by measuring two $v(H_x)$s under the influence of different out-of-magnetic field biases, $H_{z1}$ and $H_{z2}$[24]. The ratio $R_{12}$ of these two $v(H_x)$s is then written as

$$\ln[R_{12}(H_x)] = -\alpha(H_x)\left(H_{z1}^{-1/4} - H_{z2}^{-1/4}\right), \tag{1}$$

which contains only the symmetric contribution from $\alpha(H_x)$. Figure 2e plots $R_{12}(H_x)$ with $H_{z1}$ =2.1 mT and $H_{z2}$ =1.9 mT. The black curve in the plot is the best fit to guide the symmetry of $R_{12}$. The equations and parameters for the best fit will be discussed later. The figure clearly shows that the inversion symmetry axis of $R_{12}(H_x)$ becomes identical to that (blue dotted line) of the flow regime. This observation therefore supports the claims that the inversion symmetry axis of the flow regime corresponds to the true $H_{\mathrm{DMI}}$ and the present analysis method is valid to extract the symmetric contribution in the creep regime. Therefore, one can conclude that $H_{\mathrm{DMI}}$ can be uniquely determined for both creep and flow regimes. Hereafter, we will denote the uniquely determined $H_{\mathrm{DMI}}$ as $H_{\mathrm{DMI}}^*$.

Similarly, $v_0(H_x)$ can be determined by using the relation

$$v_0(H_x) = v_1(H_x)/[R_{12}(H_x)]^\gamma, \tag{2}$$

where $\gamma \equiv H_{z1}^{-1/4}/(H_{z1}^{-1/4} - H_{z2}^{-1/4})$ and $v_1$ is the DW speed measured under $H_{z1}$. Figure 2f shows $v_0(H_x)$ determined with $H_{z1} = 2.1$ mT and $H_{z2} = 1.9$ mT. Though the data are slightly scattered since the statistical error in $R_{12}$ is amplified greatly because of a large $\gamma$, the plot exhibits a noticeable variation in $v_0(H_x)$, as indicated by the solid line of the best linear fit. Such sizeable variation in $v_0(H_x)$ verifies that the antisymmetric contribution of the DW speed is mainly attributed to $v_0(H_x)$, since $\alpha(H_x)$ is solely responsible for the symmetric contribution. Such antisymmetric variation of $v_0(H_x)$ can be caused by several reasons such as chiral damping[18], asymmetric DW width[25], or $H_x$-induced magnetization tilting inside the domains adjacent to the DWs. Presently, we are unable to distinguish these possibilities and further deliberate measurements are desired for future studies.

By use of $H_{\text{DMI}}^*$, the antisymmetric contributions can be further analyzed. Figure 3(a) plots the asymmetry $A$ of the DW speed with respect to $\Delta H_x$ (i.e. $\Delta H_x \equiv H_x + H_{\text{DMI}}^*$) for the creep (red) and flow (blue) regimes, where

$$A(\Delta H_x) \equiv \frac{v(-H_{\text{DMI}}^* + \Delta H_x) - v(-H_{\text{DMI}}^* - \Delta H_x)}{v(-H_{\text{DMI}}^* + \Delta H_x) + v(-H_{\text{DMI}}^* - \Delta H_x)}. \tag{3}$$

It is clear from the figure that the creep regime exhibits a large asymmetry, in contrast with zero asymmetry in the flow regime. In the creep regime, one can readily derive the relation $A(\Delta H_x) \cong \beta \Delta H_x$ within the context of the creep criticality, where

$$\beta \equiv \frac{dv_0/d\Delta H_x|_{\Delta H_x=0}}{v_0(-H_{\text{DMI}}^*)}. \tag{4}$$

The value of $\beta$ (= 4.6±0.1 T$^{-1}$) determined from the red symbols of Fig. 3a exactly coincide with the value (= 4.2±1.4 T$^{-1}$) determined from Fig. 2f within the experimental accuracy. The exact conformity again supports the validity of our approach.

Figure 3b summarizes the experimentally determined $\beta$ (black) and $H_\mathrm{min}$ (green) with respect to $H_z$. It is evident from the figure that there exist sizeable asymmetries in the creep regime with a small $H_z$, but the asymmetry quickly decays as $H_z$ increases in the flow regime. Similarly, $H_\mathrm{min}$ approaches $-H^*_\mathrm{DMI}$ as $H_z$ increases. The dotted lines indicate a simple exponential decay function.

According to ref.14, $\alpha(H_x) = \alpha_0[\sigma_\mathrm{DW}(H_x)/\sigma_0]^{1/4}$, where $\sigma_\mathrm{DW}$ is the DW energy density, $\sigma_0$ is the Bloch-type DW energy density, and $\alpha_0$ is a scaling constant. Recent studies[14,15] on the DMI effect on DWs have revealed that $\sigma_\mathrm{DW}(H_x)$ is given by

$$\sigma_\mathrm{DW}(H_x) = \begin{cases} \sigma_0 - 2\lambda K_\mathrm{D} \left|\dfrac{H_x + H^*_\mathrm{DMI}}{H_\mathrm{K}}\right|^2 & \text{for } |H_x + H^*_\mathrm{DMI}| \leq H_\mathrm{K} \\ \sigma_0 + 2\lambda K_\mathrm{D} - 4\lambda K_\mathrm{D} \left|\dfrac{H_x + H^*_\mathrm{DMI}}{H_\mathrm{K}}\right| & \text{otherwise} \end{cases}. \quad (5)$$

Here, $K_\mathrm{D}$ is the DW anisotropy energy density, $\lambda$ is the DW width, and $H_\mathrm{K}$ ($\equiv 4K_\mathrm{D}/\pi M_\mathrm{S}$) is the DW anisotropy field. The solid line in Fig. 2e is the best fit by these equations with the following best fitting parameters: $\alpha_0$ = 3.81 T$^{1/4}$, $\lambda K_\mathrm{D}/\sigma_0$ = 0.027, $H_\mathrm{K}$ = 29.7 mT, and $H^*_\mathrm{DMI}$ = 59.3 mT. For the case in which $v_0$ exhibits a finite asymmetry i.e. $v_0(-H^*_\mathrm{DMI} + \Delta H_x) \cong v_0(-H^*_\mathrm{DMI}) + \beta \Delta H_x$, one can again easily calculate that the apparent minimum $H_\mathrm{min}$ can be written as

$$H_\mathrm{min} \cong -H^*_\mathrm{DMI} - \eta\beta H_z^{1/4}, \quad (6)$$

where $\eta \equiv \sigma_0 H_\mathrm{K}^2/\alpha_0 \lambda K_\mathrm{D}$. The prediction given by Eq. (6) is experimentally confirmed as observed by the linear relation between $\beta H_z^{1/4}$ and $H_{\min}$ in Fig. 3c. Therefore, one can conclude that the asymmetry is the origin of the deviation in $H_{\min}$ and therefore, $H_{\min}$ converges to $-H_{\mathrm{DMI}}^*$ as the asymmetry vanishes in the flow regime.

The DW motion in the flow regime can be described by the 1-dimensional DW model[26,27] based on the Landau-Lifshitz-Gilbert (LLG) equation[28]. It is well known that, under a $H_z$ larger than the Walker breakdown field[29], the DW exhibits precessional motion[30]. By solving the 1-dimensional DW model, the DW speed is given by $v = \lambda(\gamma_0 H_z - 2\pi/T)/\alpha_G$, where $\alpha_G$ is the Gilbert damping constant and $\gamma_0$ is the gyromagnetic ratio. The precession period $T$ is then written as

$$T = \frac{1+\alpha_G^2}{\gamma_0} \int_0^{2\pi} \frac{d\psi}{H_z - \alpha_G \frac{\pi}{2}(H_x + H_{\mathrm{DMI}})\sin\psi + \alpha_G \frac{\pi}{2} H_K \sin\psi \cos\psi}, \tag{7}$$

where $\psi$ is the angle of the magnetization inside the DW. Since $H_z \gg \alpha_G H_K$ for the experimental condition in the flow regime with a small $\alpha_G$, it is a good approximation to write $T$ as

$$\begin{aligned} T &\approx \frac{1+\alpha_G^2}{\gamma_0} \int_0^{2\pi} \frac{d\psi}{H_z - \alpha_G \frac{\pi}{2}(H_x + H_{\mathrm{DMI}})\sin\psi} \\ &= \frac{1+\alpha_G^2}{\gamma_0} \frac{2\pi}{\sqrt{H_z^2 - \alpha_G^2 \left(\frac{\pi}{2}\right)^2 (H_x + H_{\mathrm{DMI}})^2}}. \end{aligned} \tag{8}$$

Figure 4 plots the numerical evaluation $T/T_{\min}$ of Eqs. (7) and (8) for our experimental condition, where $T_{\min}$ is defined as $T$ at $H_x = -H_{\mathrm{DMI}}$. The micromagnetic simulation results by use of the object oriented micromagnetic framework (OOMMF) are plotted together. The figure clearly shows that all results match each other with accuracy better than

1%. Therefore, it is good to write $v$ as

$$v \approx \frac{\lambda \gamma_0}{\alpha_G} \left( H_z - \frac{1}{1+\alpha_G^2} \sqrt{H_z^2 - \alpha_G^2 \left(\frac{\pi}{2}\right)^2 (H_x + H_{\mathrm{DMI}})^2} \right)$$

$$\approx \frac{\lambda \gamma_0 \alpha_G}{1+\alpha_G^2} H_z \left( 1 + \left(\frac{\pi}{2}\right)^2 \frac{(H_x + H_{\mathrm{DMI}})^2}{2H_z^2} \right), \tag{9}$$

which exhibits symmetric behavior with a parabolic dependence on $H_x + H_{\mathrm{DMI}}$. Note that the parabolic variation originates from the suppression of precessional DW motion because of the additional energy barrier enhanced by an in-plane magnetic field[13]. It is also worthwhile to note that $\lambda$ also varies with respect to $H_x$[25], but the variation of $\lambda$ is expected to be less than a few tens of percent. Therefore, the large variation of $v$ observed in Fig. 2d is mostly attributed to the suppression of the precessional DW motion rather than $\lambda$ variation.

Finally, we would like to mention the possible antisymmetric contributions in the flow regime. Yoshimura et al.[19] recently demonstrated that above the Walker breakdown, the DMI generates soliton-like DWs partitioned by vertical Bloch lines. These DWs stay in the Bloch type configuration without precession; therefore, all the effects related to the DW chirality such as the chiral damping[18] and chiral $\lambda$ variation[25] are expected to disappear. In addition, Kim et al.[25] have shown that the Bloch-type DWs have the same DW width and DW energy density irrespective of the strength of $H_x$. The $H_x$-induced magnetization tilting possibly causes additional variation in $\lambda$. However, because of the large anisotropy field (~1 T) in typical PMA films, the variation of $\lambda$ is estimated to be small (< 3%) within the range of $H_x$. Such a small variation of $\lambda$ can cause small asymmetry (< 5%), which causes a deviation of approximately 1.8 mT in $H_{\mathrm{min}}$ that is negligibly small in comparison with other experimental inaccuracies.

In conclusion, we examine the nature of the asymmetric behavior in DW motion over the creep and flow regimes. Based on the distinct dependence of the DW speed on the in-plane and out-of-plane magnetic fields, the symmetric and antisymmetric contributions of the DW speed are decomposed, enabling one to quantify the pure effect of the DMI. The results show that the antisymmetric contribution vanishes gradually across the regimes, while the symmetric contribution remains unchanged, confirming the uniqueness of the DMI-induced magnetic field across the regimes. The present observation elucidates the underlying physics on the recent puzzling issue in the DMI-related chiral DW dynamics.

**References**

1. Siegel, J. S. Single-handed cooperation. Nature **409**, 777-778 (2001).

2. Ellis, J. Antimatter matters. Nature **424**, 631-634 (2003).

3. Bode, M. *et al*. Chiral magnetic order at surfaces driven by inversion asymmetry. Nature **447**, 190-193 (2006).

4. Yu, X. Z., Onose, Kanazawa, Y. N., Park, J. H., Han, J. H., Matsui, Y., Nagaosa, N. & Tokura, Y. Real-space observation of a two-dimensional skyrmion crystal. Nature **465**, 901-904 (2010).

5. Dzialoshinskii, I. E. Thermodynamic theory of weak ferromagnetism in antiferromagnetic substances. Sov. Phys. JETP **5**, 1259-1272 (1957).

6. Moriya, T. Anisotropic superexchange interaction and weak ferromagnetism. Phys. Rev. **120**, 91-98 (1960).

7. Chen, G. *et al*. Novel chiral magnetic domain wall structure in Fe/Ni/Cu(001) films. Phys. Rev. Lett. **110**, 177204 (2013).

8. Ryu, K.-S., Thomas, L., Yang, S.-H. & Parkin, S. Chiral spin torque at magnetic domain walls. Nature Nanotech. **8**, 527-533 (2013).

9. Haazen, P. P. J., Murè, E., Franken, J. H., Lavrijsen, R., Swagten, H. J. M. & Koopmans, B. Domain wall depinning governed by the spin Hall effect. Nature Mater. **12**, 299-303 (2013).

10. Emori, S., Bauer, U., Ahn, S.-M., Martinez, E. & Beach, G. S. D. Current-driven dynamics of chiral ferromagnetic domain walls. Nature Mater. **12**, 611-616 (2013).

11. Yang, S.-H., Ryu, K.-S. & Parkin, S. Domain-wall velocities of up to 750 m s−1 driven by exchange-coupling torque in synthetic antiferromagnets. Nature Nanotech. **10**, 221-226 (2015).

12. Moon, K.-W. *et al*. Magnetic bubblecade memory based on chiral domain walls. Sci. Rep. **5**, 9166 (2015).

13. Thiaville, A., Rohart, S., Jué, É., Cros, V. & Fert, A. Dynamics of Dzyaloshinskii domain walls in ultrathin magnetic films. Europhys. Lett. **100**, 57002 (2012).

14. Je, S.-G. *et al*. Asymmetric magnetic domain-wall motion by the Dzyaloshinskii-Moriya interaction. Phys. Rev. B **88**, 214401 (2013).

15. Hrabec, A. *et al*. Measuring and tailoring the Dzyaloshinskii-Moriya interaction in perpendicularly magnetized thin films. Phys. Rev. B **90**, 020402(R) (2014).

16. Pizzini, S. *et al*. Chirality-induced asymmetric magnetic nucleation in Pt/Co/AlOx ultrathin microstructures. Phys. Rev. Lett. **113**, 047203 (2014).

17. Vaňatka, M. *et al*. Velocity asymmetry of Dzyaloshinskii domain walls in the creep and flow regimes. J. Phys.: Condens. Matter **27**, 326002 (2015).

18. Jué, E. *et al*. Chiral damping of magnetic domain walls. Nature Mater. **15**, 272-277 (2016).

19. Yoshimura, Y. *et al*. Soliton-like magnetic domain wall motion induced by the interfacial Dzyaloshinskii–Moriya interaction. Nature Phys. **12**, 157-162 (2016).

20. Kim, D.-H. *et al*. Maximizing domain-wall speed via magnetic anisotropy adjustment in Pt/Co/Pt films. Appl. Phys. Lett. **104**, 142410 (2014).

21. Kim, K.-J. *et al*. Interdimensional universality of dynamic interfaces. Nature **458**, 740-742 (2009).

22. Metaxas, P. J. *et al*. Creep and flow regimes of magnetic domain-wall motion in ultrathin Pt/Co/Pt films with perpendicular anisotropy. Phys. Rev. Lett. **99**, 217208 (2007).

23. Lemerle, S. *et al*. Domain wall creep in an Ising ultrathin magnetic film. Phys. Rev. Lett. **80**, 849-852 (1998).

24. Kim, D.-Y., Kim, D.-H., Moon, J. & Choe, S.-B. Determination of magnetic domain-wall types using Dzyaloshinskii–Moriya-interactioninduced domain patterns. Appl. Phys. Lett. **106**, 262403 (2015).

25. Kim, D.-Y., Kim, D.-H. & Choe, S.-B. Intrinsic asymmetry in chiral domain walls due to the Dzyaloshinskii–Moriya interaction. Appl. Phys. Express **9**, 053001 (2016).

26. Slonczewski, J. C. AIP Conf. Proc. **5**, 170-174 (1972).

27. Hillebrands, B. & Thiaville, A. Spin Dynamics in Confined Magnetic Structures III, Topics in Applied Physics **101**, 161-205 (2006).

28. Gilbert, T. L. A phenomenological theory of damping in ferromagnetic materials. IEEE Trans. Magn. **40**, 3443-3449 (2004).

29. Beach, G. S. D., Nistor, C., Knutson, C., Tsoi, M. & Erskine, J. L. Dynamics of field-driven domain-wall propagation in ferromagnetic nanowires. Nature Mater. **4**, 741-744 (2005).

30. Hayashi, M., Thomas, L., Rettner, C., Moriya, R. & Parkin, S. S. P. Direct observation of the coherent precession of magnetic domain walls propagating along permalloy nanowires. Nature Phys. **3**, 21-25 (2007).

**Figure Captions**

Figure 1. **Plot of $v$ as a function of $H_z$ with $H_x = 0$. a**, Linear scale plot for the flow regime and **b**, creep scale plot for the creep regime. Both the blue and red lines show the best linear fit.

Figure 2. **Plot of $v/v_{\min}$ with respect to $H_x$ under various strengths of $H_z$: a**, 1.9 **b**, 3.4 **c**, 17, and **d**, 41 mT. In each plot, the solid line shows the best parabolic fit with Eq. (9) and the purple arrow indicates $H_{\min}$ for $v_{\min}$. **e**, Plot of $R_{12}$ with respect to $H_x$ with $H_{z1} = 2.1$ mT and $H_{z2} = 1.9$ mT. The solid line shows the best fit with Eq. (5). **f**, Plot of $v_0$ with respect to $H_x$. The solid line shows the best linear fit.

Figure 3. **Asymmetry of the DW speed over the creep and flow regimes. a**, Plot of $A$ with respect to $\Delta H_x$ for the creep (red) and flow (blue) regimes. The solid lines show the best linear fit. **b**, Plot of $\beta$ (black) and $H_{\min}$ (green) with respect to $H_z$. The dotted lines indicate a simple exponential decay function. **c**, Plot of $\beta H_z^{1/4}$ with respect to $H_{\min}$. The black solid line shows the best linear fit.

Figure 4 **Plot of $T/T_{\min}$ with respect to $H_x$**, calculated by Eq. (7) (solid line) and Eq. (8) (circular symbols) as well as by micromagnetic simulation (square symbols) at $H_z = 51$ mT.

**Methods**

Methods and any associated references are available in the online version of the paper.

**Acknowledgements**

This work was supported by a National Research Foundations of Korea (NRF) grant that was funded by the Ministry of Science, ICT and Future Planning of Korea (MSIP) (2015R1A2A1A05001698 and 2015M3D1A1070465). D.-H.K. was supported by a grant funded by the Korean Magnetics Society. B.-C.M. was supported by the KIST institutional program and Pioneer Research Center Program of MSIP/NRF (2011-0027905).

**Author contributions**

D.-H.K. planned and designed the experiment and S.-B.C. supervised the study. D.-H.K. and D.-Y.K. carried out the measurement. S.-C.Y. and B.-C.M. prepared the samples. S.-B.C. and D.-H.K. performed the analysis and wrote the manuscript. All authors discussed the results and commented on the manuscript.

**Additional information**

Correspondence and request for materials should be addressed to S.-B.C.

**Competing financial interests**

The authors declare no competing financial interests.

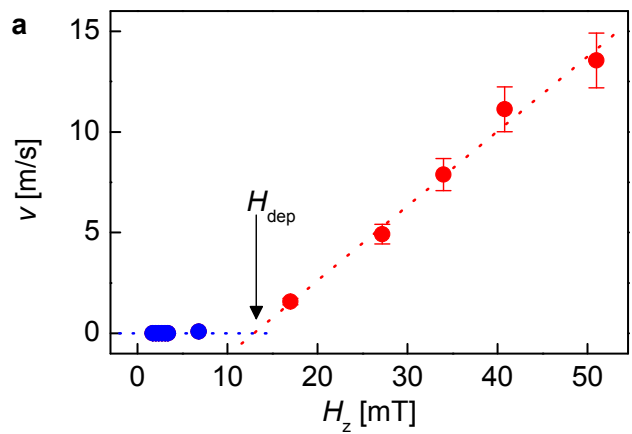

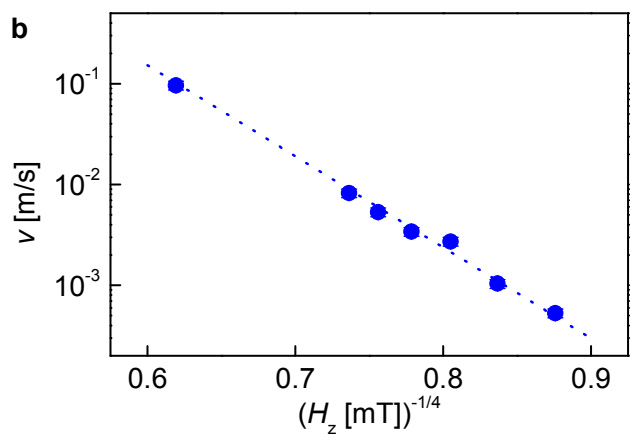

Figure 1

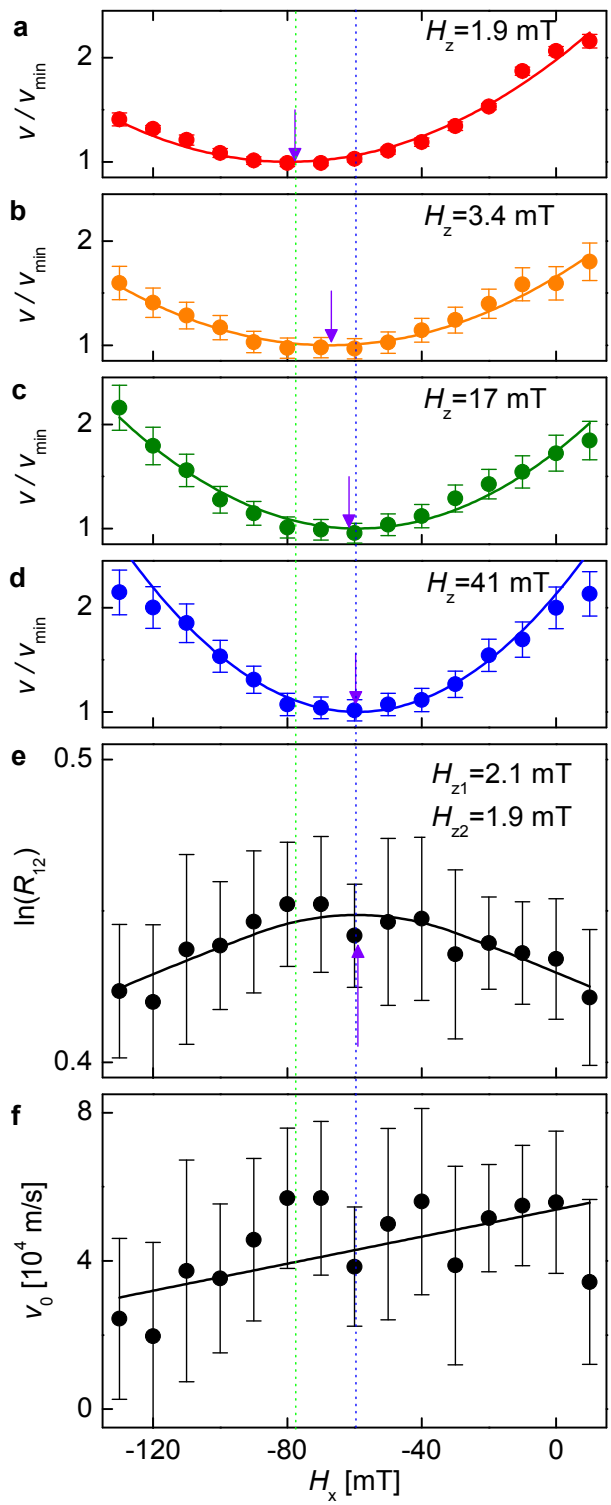

Figure 2

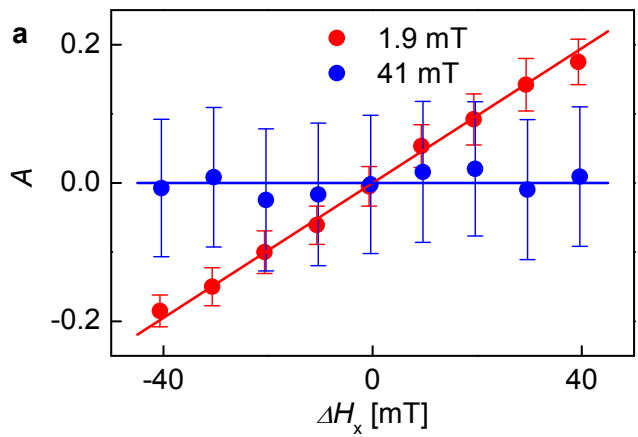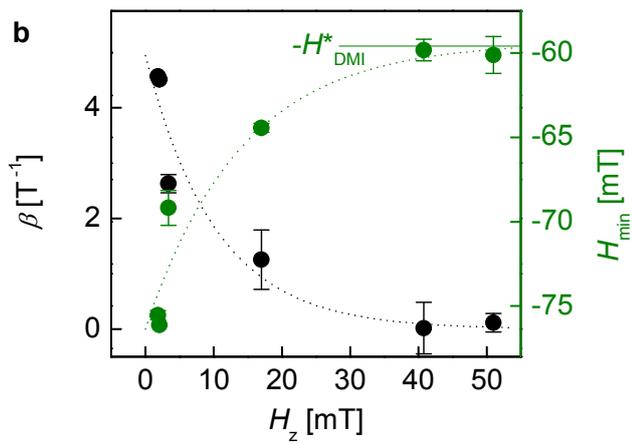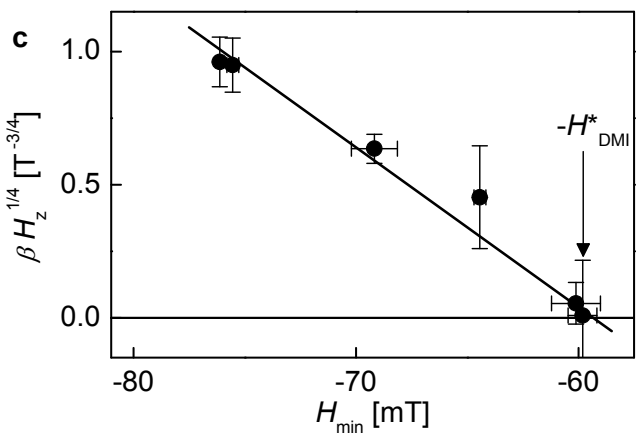

Figure 3

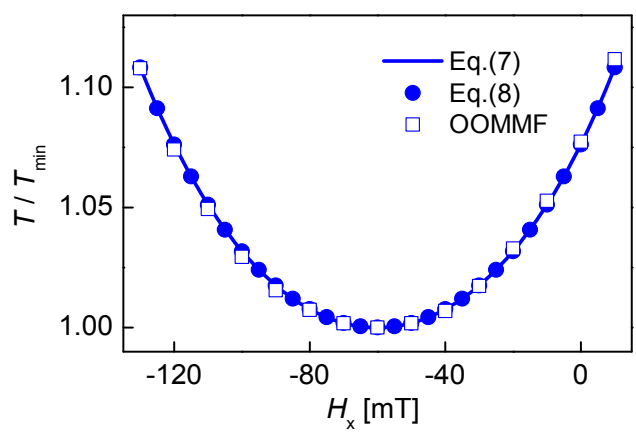

Figure 4